# Research of Load Testing and Result Based on Loadrunner

**Pratibha Fageria[1], Dr. Manju Kaushik[2]**

[1](M. Tech, Department CSE, JECRC University, Jaipur, India)
[2](Associate Professor, Department CSE, JECRC University, Jaipur, India)

**ABSTRACT :** In this paper, we made the plan of a load testing, and got results by means of the LoadRunner which is an automatic load testing tool. We combined with the characteristics of electronic commerce system and did the load testing and analysis the result of load test by means of the LoadRunner. We fully described the characteristics of the electronic commerce application, designed the reasonable test cases, and simulated the practical scenario. In the process of running Load Runner, we arranged the appropriate transactions and rendezvous, and designed the truthful test network environment. The plan was applied to the load testing phase of the telecommunication equipment sales system of special products. We analyzed the load testing results, proposed the improving measures, and realized the optimization of the telecommunication equipment sales system and also found the defect of the system when the massive users access the system and guided the system improvement using the test result.

## I. INTRODUCTION

The aim of software testing process is to identify all the defects existing in a software product. It is the process of exercising and evaluating a system or system components by manual automatic means to verify that it satisfies specified requirements or to identify differences between expected and actual results. There are two ways of testing that are manual or automation. Manual testing [1] carried out by the testers. Testers test the software manually for the defects. It requires a tester to play the role of an end user, and use most of all features of the application to ensure its correct behavior. They follow a written test plan that leads them through a set of important test cases. The problems with manual testing are, it is very time consuming process, not reusable, has no scripting facility, great effort required, and some errors remain uncovered. **Load testing [2]** is the process of putting demand on a system or device and measuring its response. Load testing is performed to determine a system's behavior under both normal and anticipated peak load conditions. It helps to identify the maximum operating capacity of an application as well as any bottlenecks and determine which element is causing degradation. When the load placed on the system is raised beyond normal usage patterns, in order to test the system's response at unusually high or peak loads, it is known as stress testing. The load is usually so great that error conditions are the expected result, although no clear boundary exists when an activity ceases to be a load test and becomes a stress test. **HP LoadRunner software** allows you to prevent application performance problems by detecting bottlenecks before a new system or upgrade is deployed. The testing solution Load Runner enables you to test rich Internet applications, Web 2.0 technologies, ERP and CRM applications, and legacy applications. It gives you a picture of end-to-end system performance before going live so that you can verify that new or upgraded applications meet performance requirements. And it reduces hardware and software costs by accurately predicting application scalability and capacity. [3]. this paper is structured as follow.

In section 2 give ideas about the load testing and its tool (Load Runner). Section 3 presents the implementation of load testing. Section 4 gives results and discussion. Conclusion and future work are summarized in section 5.To bridge the gap between existing web testing techniques and main new feature provided by web application. The server side can be tested using any conventional testing technique. Client side testing can be performed at various levels. The se-lenium tool is very popular capture-replay tool and allows DOM based testing by capturing user session i.e. events fired by user. Such tool can access the DOM and shows expected UI behaviour and replay the user session. So today's need is a testing tool which can test user session and generate test cases on the basis of expected UI behaviour as per event fired by user.

**State Based Testing:** Marchetto proposed a state based testing technique [20, 21]. Idea is that the states of client side components of an AJAX application need to be taken into account during testing phase [21]. State based testing technique for AJAX is based on the analysis of all the states that can be reached by the client-side pages of the application during its execution. Using AJAX, HTML elements—like TEXTAREA, FORM, INPUT, A, LI, SELECT, OL, UL, DIV, SPAN,

etc.—can be changed at runtime according to the user interactions. In this testing the HTML elements of a client-side page character-size the state of an AJAX Web page, and their corresponding values are used for building its finite state model. State based technique results indicate that state based testing is powerful and can reveal faults otherwise unnoticed or very hard to detect using existing techniques.

## II. LOAD TESTING AND ITS TOOL

### Why Load Testing?
In today's world of e-business, your customers and business partners want their Web sites to be competitive. The criteria for a competitive Website are the following: pages that download immediately and efficient and accurate online transactions. Online consumer and B2B marketplaces are more and more competitive. Companies cannot must be very careful and check the ability of their Web-based applications to accommodate multiple, simultaneous users who are connecting themselves to a Web site or engaging in online transactions. Those enterprises appeal more and more to the purchase of an application load-testing tool. Load testing is the processes that set the service request web client number in the input, and gradually increase the client number of the web service request. The tester can get the client average response time, and compare the average response time after the client number is increased every time. The load testing of web applications will be able to evaluate the operation of all parts of the Web server, including the CPU, memory, process, hard disk response time, etc. Load testing can also evaluate the network movement situation, and monitor the cause of the delay caused by the network.

### What is a load testing tool?
The principle of a load testing tool is to simulate the behavior of real users with "virtual" users. The load-testing tool can then record the behavior of the site under the load and give information on the virtual users' experiences. Load-testing software is often distributive in nature. It is applied on multiple servers running simultaneously, with each server simulating multiple virtual users. In many cases, the testing company has developed its own proprietary browser that can be combined with a set of instructions tailored to the testing of each client business. The testing company maintains ongoing records of the virtual users' experiences at the test site, including response times and errors.

**Load testing tool (Load Runner): -** Load Runner is a kind of load testing tools that can forecast the behavior and properties of the system. Load Runner find and make sure the problem, through the simulation of millions of users implement concurrent load and real-time performance monitoring. Through the use of Load Runner, enterprise can shorten test time in maximum limit, optimize performance, and speed up the release cycle of the application system. The main steps of the load testing by means of Load Runner include: making load test plan, test scripts, create running scene, the running of load testing, monitoring scene, the analysis of results. Load Runner contains many components, some important common are Visual User Generator (here in after referred to as VuGen), Controller, Analysis. Load Runner advantages include: testers can create virtual users by means of the VuGen easily and set up test scripts; testers can quickly build the test plan of multiple virtual users by meaning of Controller; testers can observe the performance of the application system by means of the Load Runner real-time monitor; testers can quickly find the position and the reason of the system's error by means of Analysis tools, and make the corresponding modification.

## III. IMPLEMENTATION OF LOAD TESTING

### 1. Making Load Test Plan

**1.1. Load testing goal: -** This paper implemented load testing for the e-commerce applications system by means of Load Runner, analyzed results, and improved the performance of the system based on the test results.

**1.1.1 Design Test Case: -** When testers design load test cases, the testers should take the application characteristics of the electronic commerce system into consideration.

**1.1.2 The construction of the load testing environment: -** The load testing object of the telecommunication equipment sales system will be deployed on the server, so the network environment of the load testing simulated the practical environment of the running system. Considering the characteristics of the server's network environment, we built the open environment for load testing, instead of the local area network environment.

## 1.2 Build Test Scripts

**1.2.1 Record test scripts: -** According to the above test cases, we recorded three test scripts by means of the Visual User Generator which is a component of the Load Runner. The three test scripts are the script of browsing commodity, the script of searching commodity, and the script of online shopping.

**1.2.2 The development of affairs:-** The usability of the basic script is generally not useful; tests need deploy affairs perfect script in the basic script, in order to analyze the response time of the typical operation, and to increase the script availability. There are two ways of the deploying affairs in the Script: first, during record the test script, testers will insert the Start Transaction of the affairs in the front of the sequence of operations, and insert the End Transaction of the affairs in the rear of the sequence of operations by means of the toolbar; Second, after record the script, testers insert the start and end points of the affairs in proper position according to the script meaning.

**1.3 The development of set points: -** The purpose of the deployment of the set-points in the test script is to measure the server performance index under the big pressure.

## 2. Create running Scene

Before running the load testing, in order to simulate the actual operation environment preferably, testers must create a reasonable operation scene by considering the reality.

## IV. RESULT AND DISCUSSION

After starting the load testing, Load Runner monitors the operation situation of the servers, through adding performance counters in the operation scene. Load Runner can monitor the Web server, database server, network delay time operation indicators. The indicators monitored by Load Runner include: Available Mbytes, Page/sec, etc.

## V. CONCLUSION AND FUTURE SCOPE

The main motive behind this paper to give a idea of load testing and load testing tool (Load Runner). The load testing of the B/S application system is the necessary test process before releasing the system, and can find the bottlenecks of the Web application system under pressure. As more and more web technologies have moved a long way to create web application. Web testing plays an important role. Here in this paper we discussed two well known testing techniques state based testing and invariant based testing. While these approaches are tested successful on various case studies, many problem remains, related to mainly scalability issue. How to capture user session data? How to avoid state explosions problem or how to reduce state spaces? How to improve FSM recovery steps, in order to automatically infer user session based test cases. In this research work, DOM manipulation of code into an FSM needs a proper technique. The experiments conducted in this direction are able to generate test case for semantically interacting events and proofs are available that long sequences generates huge test cases and having higher fault exposing capability [8].Future work can be to reduce state space reduction by applying any path seeding algorithm for automatically generating FSM.

In this paper, we have covered resemblance and differences between web application testing and traditional software testing. We considered web testing with respect to various web testing techniques and Web testing tools. This research paper is providing help to get information about existing web testing technique, current scenario of web testing and proposing new research direction in web testing field. The main conclusion is that all testing are fully dependent on implementation technologies and future testing techniques have to adapt hetero-generous and dynamic nature of web application. This finding remarks that, there is a need to generate a test environment to test latest web technology designed web application and exercise each of them. New testing issues can arise for testing web services for improving effectiveness and efficiency of web. The load test plan and the application method are versatile, and widely value. In future we extend our research into ecommerce web applications and enterprises utility. This paper also gives idea about other load testing tools.